\documentclass[12pt]{article}
\usepackage{subeqn}
\usepackage{fullpage}

\newcommand\pr{\prime}
\newcommand\be{\begin{equation}}
\newcommand\ee{\end{equation}}
\newcommand\bea{\begin{eqnarray}}
\newcommand\eea{\end{eqnarray}}
\newcommand\nn{\nonumber}

\newcommand\bdm{\begin{displaymath}}
\newcommand\edm{\end{displaymath}}

\def\pmb#1{\setbox0=\hbox{#1}%
  \kern-.025em\copy0\kern-\wd0
   \kern.05em\copy0\kern-\wd0
   \kern-.025em\raise.0433em\box0 }
\def\bx{\mbox{\boldmath$x$}}
\def\by{\mbox{\boldmath$y$}}

\def\bWW{\mbox{\boldmath$W$}}

\def\bPP{\mbox{\boldmath$P$}}
\def\bb{\mbox{\boldmath$b$}}
\def\rd{\mathrm d}
\newcommand\bnab{\mbox{\boldmath$\nabla$}}

\begin{document}

\title{\large {\bf  Momentum in stochastic quantum mechanics\thanks{{\em Letters in Mathematical Physics\/} {\bf 5}  (1981) 523--529. {\em Copyright\/} \copyright \,1981 {\em D.\ Reidel Publishing Company.}}}}

\author{{\normalsize Mark Davidson}\thanks{
Current Address: Spectel Research Corporation, 807 Rorke Way, Palo Alto, CA   94303 
\newline  Email:  mdavid@spectelresearch.com, Web: www.spectelresearch.com}\\
\normalsize{\em Department of Physics, San Jose State University, San Jose, CA  95192, U.S.A.}}

\date{}

\maketitle
\begin{abstract} Momentum is analyzed as a random variable in stochastic quantum mechanics.  Arbitrary potential energy functions are considered.  The oscillator is presented as an example.
\end{abstract}

\section{Introduction}

In stochastic quantum mechanics a particle's position is a time-dependent random variable which traverses a continuous path in space, its motion being described by a Markov process \cite{nelson1,nelson2}.  The position plays a more fundamental role than the momentum or than any other observable in this theory.  This elevated role of position is in sharp contrast to the usual formalism of quantum mechanics where coordinates and momenta are on the same footing, both being represented by operators on a Hilbert space.  It is desirable to see if random variables can be associated with momenta as well as coordinates in stochastic quantum mechanics.  For the case of the free particle, this problem has been solved by Shucker \cite{shucker1,shucker2}.  We present here a generalization of Shucker's work to the case of arbitrary potential.

Random variables are defined here whose distributions are the same as those of the corresponding quantum mechanical operators.  For the ground state of the oscillator they are explicitly calculated as an example.

It would be interesting to find an algorithm for constructing a random variable whose distribution is the same as that of a general quantum mechanical observable, in general a function of both coordinates and momenta.  It is not known whether such a program can be carried out without enlarging the probability space of stochastic quantum mechanics.  The differential space theory of Wiener and Siegel \cite{wiener/siegel} provides random variables for all Hermitian operators, but the probability space in that theory seems to be larger than the space of sample paths in stochastic quantum mechanics.  In any event, the correspondence rule between ransom variables and Hermitian operators must be non-linear in order to avoid conflict with Von Neumann's famous theorem on hidden variables [6, p.\ 209], 
a point made by Wiener and Siegel [5, p.\ 148] and also by Bell \cite{bell}.  We do not conflict with Von Neumann's theorem here since we do not assert that the sum of two non-commuting operators should be represented by the corresponding sum of their random variables.

The random variable for momentum which is presented here is quite complicated except for the case of a free particle.  We find that in general it cannot be determined analytically and that it depends upon the state of the quantum mechanical system.  Nevertheless, its existence is guaranteed in a wide range of cases, and it may yield some insight into the stochastic interpretation of quantum mechanics.

The results presented are derived for the generalized F\'enyes--Nelson model \cite{davidson1}--\cite{davidson3} where the diffusion parameter is arbitrary.  Various other applications of stochastic quantum mechanics are discussed in \cite{yasue}--\cite{guerra2}.

\section{Momentum as a random variable}

Consider Schr\"odinger's equation
\begin{equation}
\left[ -\frac{1}{2} \Delta +V(x) \right]\psi =i \frac{\partial}{\partial t} \psi
\label{1}
\end{equation}
where $\Delta$ is the $N$-dimensional Laplacian, $x$ is an $N$-dimensional coordinate, and $V$ a potential function.  Units have been chosen to make the coefficient of $\Delta$ equal to $-\frac{1}{2}$.  Suppose that at time $t_0$ the wave function is given by
\begin{equation}
\psi (x, t_0) =\psi_0 (x).
\label{2}
\end{equation}
Then Schr\"odinger's equation may be integrated, provided $\psi_0$ is sufficiently regular.

A stochastic model of Schr\"odinger's equation is constructed by defining $R$ and $S$ by
\begin{equation}
\psi=e^{R+iS}
\label{3}
\end{equation}
and identifying $x$ with the stochastic  process defined by the following stochastic differential equation
\begin{equation}
d\bx =\bb (x, t) {\rm d} t +{\rm d}\bWW
\label{4}
\end{equation}
where 
\begin{equation}
\bb =2\nu \bnab R+\bnab S
\label{5}
\end{equation} and where the diffusion parameter $\nu$ is defined by
\begin{equation}
E\left({\rm d} W_i {\rm d}W_j\right) =2\nu \delta_{ij} \, {\rm d} t.
\label{6}
\end{equation}
In the most general version of stochastic quantum mechanics \cite{davidson1}--\cite{davidson3}, this diffusion parameter is arbitrary. For Nelson's original theory \cite{nelson1,nelson2} $\nu$ in (\ref{6}) must be $\frac{1}{2}$.

Now, consider a free particle equation with the same starting conditions (\ref{2}):
\begin{equation}
-\frac{1}{2} \Delta \psi_F =i \frac{\partial}{\partial t} \psi_F.
\label{7}
\end{equation}
Then this free-particle equation also admits a stochastic model defined by the equation
\begin{equation}
{\rm d}\bx_F =\bb_F \, {\rm d}t +{\rm d}W_F
\label{8}
\end{equation}
where
\begin{equation}
\psi_F =e^{R_F+iS_F} \quad \mbox{ and } \quad \bb_F =2\nu \bnab R_F +\bnab S_F.
\label{9}
\end{equation}
Shucker \cite{shucker1} has proved the following result for the free-particle Schr\"odinger equation:
\bdm
\lim\limits_{T\to\infty} \; \frac{\bx_F (t_0 +T)}{T} \,= \bPP 
\edm
exists and has a probability density given by
\begin{equation}
\rho(P) =\frac{\psi^* (P) \psi(P)}{(2\pi)^3}\, ,\quad \psi(P) =\int {\rm d}^3 x \, e^{-iP\cdot {\footnotesize \bx}} \psi (x, t_0).
\label{10}
\end{equation}
Shucker's results apply equally well to the general version of stochastic quantum mechanics as was shown in \cite{shucker2}.

Note that $\rho(P)$ is the same quantum mechanical momentum density as the original problem with a potential at time $t_0$ since the free and interacting wave functions are both the same at this time.
  
Now consider (\ref{4}) and (\ref{8}). By fiat, require that the Wiener processes in these two equations be the same.  That is, impose the condition
\begin{equation}
\bWW =\bWW_F.
\label{11}
\end{equation}
Then $\bx_F$ and $\bx$ are related by
\begin{equation}
{\rm d}\bx -\bb (x, t) {\rm d} t={\rm d} \bx_F -\bb_F (x_F, t) {\rm d} t.
\label{12}
\end{equation}
By writing (\ref{12}), as two equations
\bea
d\bx_F &=& (-\bb(y,t) +\bb_F (x_F, t)) {\rm d} t +{\rm d}\bx \label{13}\\
\rd\by &=& \rd \bx, \qquad \by(t_0) =\bx (t_0)
\label{14}
\eea
then we obtain the standard form for a multidimensional stochastic differential equation, and since the sample  paths of $x$ are the same as those of a simple Wiener process, we may solve for $\bx_F$ in terms of $\bx$ by iteration.  This application of the Picard method was pointed out for a similar problem by Klauder \cite{klauder} who also presented a perturbation method for finding a solution.  Klauder's interaction picture method may be useful for finding approximations to the momentum in cases of weak potential.  The solution to (\ref{13}) and (\ref{14}) is obtained by iterating the following equation
\be
\bx_F (t) =x(t) +\int\limits^t_{t_0} \, \left[ \bb_F (x_F, t) -\bb(x,t) \right] \,\rd t.
\label{15}
\ee
Convergence of this iteration is ensured by the contraction  mapping theorem if the following global Lipschitz condition is satisfied
\bea
&& \left|\left(\bb_F (y_1, t) -\bb (\bx_1, t)\right) -\left( \bb_F (y_2, t )- \bb (x_2, t)\right)\right|\nn\\
&&\qquad < k \left[ \left| \bx_1 -\bx_2 \right| +\left| \by_1 -\by_2\right| \right]
\label{16}
\eea
for all $x_1, x_2, y_1$, and $y_2$;  and all $t$, and where $k$ is a constant.  See for example [2, p.\ 43]. 

Note that we have by fiat imposed the condition 
\be
\bx_F(t_0) =\bx(t_0)
\label{17}
\ee
which is consistent because both processes have the same probability density at $t=t_0$.  If we let $(\Omega, \Sigma, P)$ denote the probability space for the process $\bx(t)$, and if $\omega\in \Omega$, then $\bx(\omega, t)$ is a sample trajectory.  The above analysis provides an iterative algorithm for calculating $x_F (\omega, t)$.  Thus the free particle's sample trajectory is expressed as a random variable on the probability space of the interacting particle.

Now we need only apply Shucker's theorem.  Let $x_F$ be calculated by (\ref{15}) and consider the limit 
\be
\bPP(t_0) =\lim\limits_{T\to \infty} \;\frac{\bx_F (\omega, t_0 +T)}{T}.
\label{18}
\ee Then, by Shucker's theorem \cite{shucker2}, $\bPP$ exists and has a probability density given by (\ref{10}).  Thus we have constructed a random variable for momentum whose distribution is the same as the quantum mechanical one for a system with an arbitrary potential.  We next present a simple example to illustrate this technique.

\section{The Oscillator}
Consider the ground state of a one-dimensional harmonic oscillator which satisfies the equation 
\be
\left[ -\frac{1}{2} \;\frac{\partial^2}{\partial x^2} +\frac{1}{2} x^2\right] \, \psi =E\psi.
\label{19}
\ee
Up to a normalization constant, the solution to (\ref{19}) is
\be
\exp \left[ -x^2 /2\right]
\label{20}
\ee
so that 
\be
b(x) =-2\nu x.
\label{21}
\ee
Now consider the free particle solution with initial conditions at $t=t_0$ given by (\ref{20}):
\be
\psi_F (x,t) =\exp \left[-\frac{1}{2} x^2 /(1+i(t-t_0))\right]
\label{22}
\ee
so that from (\ref{9}) we have
\begin{subequations}
\bea
R_F&=& -\frac{1}{2} x^2_F /(1+ (t-t_0)^2)\label{23a}\\
S_F&=& \frac{1}{2} x^2_F (t-t_0) /(1+(t-t_0)^2)\label{23.b}
\eea
\end{subequations}
\be
\!\!\!\!\!\!\!\!\!\!\!\!b_F = -x_F \left[ \frac{2\nu -(t-t_0)}{1+ (t-t_0)} \right].
\label{24}
\ee
Then applying (\ref{12}) yields
\be
\rd x +2 \nu x \, \rd t =\rd x_F +x_F \frac{2\nu -(t-t_0)}{1+(t-t_0)^2} \, \rd t\label{25}
\ee
which may be directly integrated to yield
\bea
x_F (t) &=&e^{-\gamma(t)} x(t_0) +e^{-\gamma(t)} \; \int\limits^{x(t)}_{x(t_0)} \, e^\gamma \rd x +\nn\\
&&\quad +2\nu e^{-\gamma(t)} \int\limits^t_{t_0} \,e^{\gamma (t^\pr)}  x(t^\pr)\,  \rd t^\pr
\label{26}
\eea
where 
\bea
\gamma(t) &=& \int\limits^t_{t_0} \, \frac{2\nu -(t^\pr -t_0)}{1+ (t^\pr -t_0)^2} \, dt^\pr\nn\\
&=& 2\nu A \tan (t-t_0) -\frac{1}{2} \ln (1+(t-t_0)^2).
\label{27}
\eea
The momentum may be calculated from (\ref{26}).  The result, after some manipulation, is
\be
P(t_0)=e^{-\nu\pi} \int\limits^\infty_{t_0} \, x(t) e^{\gamma(t)} \left( 2\nu -\frac{\partial\gamma}{\partial t}\right) \,\rd t.
\label{28}
\ee
One can check that this expression for $P$ has the appropriate density by noting first that it has a Gaussian distribution and then by calculating its variance.  Using (\ref{27}) together with the fact that 
\be
E(x(t_1) x(t_2))=\frac{1}{2} \exp \left[ -2\nu |t_1 -t_2|\right]
\label{29}
\ee
one finds that $E(P^2)=\frac{1}{2}$ as expected.

\section{Conclusion}
The technique presented here allows one to consider momentum as a random variable on an equal footing with position in the stochastic interpretation of quantum mechanics.  It is easily generalized to allow for magnetic fields also.  The momentum depends not only on the sample path in question, but also on the state of the system.  Of course, this momentum variable satisfies the Heisenberg uncertainty principle with the coordinate.  It is hoped that these results will facilitate the construction of random variables for still more general operators involving both coordinates and momenta.

\end{document}